\documentclass[11pt]{article}

\usepackage[preprint]{acl}

\usepackage{times}
\usepackage{latexsym}
\usepackage[linesnumbered, ruled, vlined]{algorithm2e} 

\usepackage{amsmath}
\usepackage{amssymb}

\usepackage[T1]{fontenc}

\usepackage[utf8]{inputenc}

\usepackage{microtype}

\usepackage{inconsolata}

\usepackage{graphicx}

\usepackage{booktabs}  
\usepackage{multirow}  
\usepackage{float}     
\usepackage{comment}

\usepackage[breakable,skins]{tcolorbox}
\newtcolorbox{promptbox}[1]{%
    breakable,
    enhanced,
    colback=white,
    colframe=green!60!black,
    arc=6pt,
    boxrule=0.4pt,
    title=#1,
    colbacktitle=green!60!black,
    coltitle=white,
    fonttitle=\bfseries\large,
}

\title{Bridge-RAG: An Abstract Bridge Tree Based Retrieval Augmented Generation Algorithm}

\author{
\textbf{Zihang Li},
\textbf{Wenjun Liu},
\textbf{Yikun Zong}, 
\textbf{Jiawen Tao},
\textbf{Siying Dai}, 
\textbf{Songcheng Ren}, \\
\textbf{Zirui Liu},
\textbf{Yuhang Wang},
\textbf{Yanbing Jiang},
\textbf{Tong Yang}\textsuperscript{$\dagger$} 
 \\ \\
\textbf{Peking University}\\
\textbf{\textsuperscript{$\dagger$}Corresponding Author} 
\vspace{5pt}
}

\begin{document}
\maketitle

\begin{abstract}
As an important paradigm for enhancing the generation quality of Large Language Models (LLMs), retrieval-augmented generation (RAG) faces the two challenges regarding retrieval accuracy and computational efficiency.
This paper presents a novel RAG framework called Bridge-RAG.
To overcome the accuracy challenge, we introduce the concept of \textit{abstract} to bridge query entities and document chunks, providing robust semantic understanding.
We organize the abstracts into a tree structure and design a multi-level retrieval strategy to ensure the inclusion of sufficient contextual information.
While this hierarchical organization substantially improves answer quality, traversing the tree to locate the abstracts that contain a query entity inevitably introduces additional retrieval overhead.
To restore retrieval efficiency, we further integrate the Cuckoo Filter in CFT-RAG~\citep{li2026cftrag}, which provides $O(1)$ entity lookup and naturally fits the entity-to-abstract pathway of our framework.
Extensive experiments show that Bridge-RAG achieves consistent accuracy improvements across all metrics and up to $1.9\times$ faster retrieval compared to structured RAG baselines.
All related code is available in the ``software'' appendix.
\end{abstract}

\section{Introduction}
\label{submission}

Retrieval-Augmented Generation (RAG) has become an important paradigm for enhancing the generation quality of Large Language Models (LLMs) by integrating external knowledge~\citep{patrick2020retrieval}.
Existing RAG frameworks differ chiefly in how the external knowledge is organized: at one end of the spectrum, simple text stores keep documents as raw passages and retrieve them by vector similarity; at the other, structured representations---knowledge graphs and entity trees---encode the relations among concepts so that retrieval can surface richer, more coherent evidence than a flat passage list.

RAG systems face two critical challenges regarding retrieval accuracy and computational efficiency. 
First, in terms of retrieval accuracy, standard RAG frameworks usually rely on vector similarity search, which tends to retrieve passages that are semantically related but contextually fragmented or noisy. This misalignment often results in the generator receiving insufficient or irrelevant evidence, inevitably propagating errors to the final response. 
Second, a severe efficiency bottleneck persists in structured RAG frameworks, especially for Tree-RAG: once trees deepen and the leaf set climbs into the millions, walking the hierarchy to pin down the entities a query refers to becomes one of the dominant costs in the response time budget.
Our empirical results in Table~\ref{tab:combined_evaluation} indicate that
the retrieval times of standard RAG frameworks like Tree-RAG and Graph-RAG are still quite long. These two challenges hinder the deployment of RAG systems in real-world scenarios that demand high precision and low-latency responses.

\begin{figure*}[h!]

    \centering
    \includegraphics[width=\textwidth]{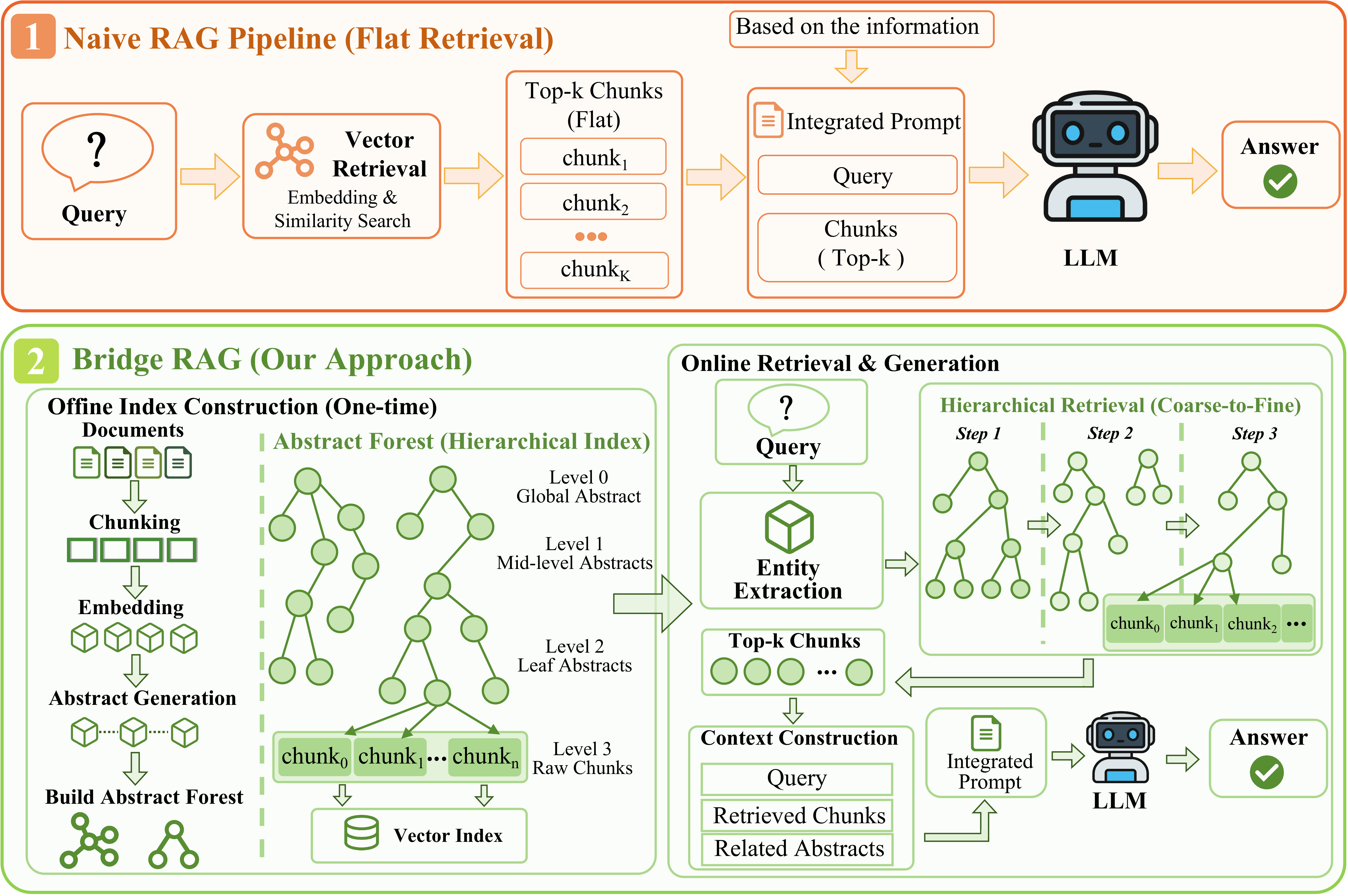}
    \caption{Comparison of Naive RAG and Bridge-RAG. (Top)~Naive RAG retrieves flat top-$k$ chunks by similarity, lacking global semantic understanding and cross-document context. (Bottom)~Bridge-RAG constructs an abstract forest offline and performs hierarchical coarse-to-fine retrieval online: query entities are located via $O(1)$ Cuckoo Filter lookup, the abstract tree is traversed for multi-level context, and fine-grained chunks are selected for prompt construction and generation.}
\label{fig:my_label0}
\end{figure*}

This paper proposes Bridge-RAG to effectively address the two challenges above.
The core innovation of Bridge-RAG is the \textit{hierarchical abstract bridge tree}: we introduce the concept of \textit{abstract} and organize abstracts into a tree structure that systematically bridges query entities and document chunks, enabling the retrieval of coherent multi-level context.
This hierarchical organization, however, increases the cost of locating relevant abstracts at retrieval time.
To restore efficiency, Bridge-RAG directly integrates the Cuckoo Filter from the recently published CFT-RAG~\citep{li2026cftrag}, which provides $O(1)$ entity lookup---significantly faster than the entity-location step in naive Tree-RAG or Graph-RAG---and naturally fits as the entry point of our entity-to-abstract pathway.
The workflow of Bridge-RAG is shown in Figure~\ref{fig:my_label0}.

Our first key design focuses on enhancing retrieval accuracy from the aspects of semantic depth and contextual breadth. 
First, we introduce the concept of \textit{abstract} to facilitate robust semantic understanding. 
Specifically, an \textit{abstract} groups several consecutive document chunks into a high-level knowledge unit. 
We organize these abstracts into tree structures where upper-level abstracts represent more general concepts and lower-level ones capture more detailed information. 
This hierarchical organization facilitates a three-tier retrieval path where \textit{abstract} serves as the bridge between \textit{query entity} and \textit{document chunk}. 
In this way, Bridge-RAG ensures the captured content is semantically relevant. 
Second, we design a multi-level retrieval strategy to ensure the inclusion of sufficient contextual information.
Specifically, Bridge-RAG systematically traces the hierarchical lineage of the abstract tree, retrieving document chunks associated with both parent and child abstracts alongside the target node.
This strategy ensures that the retrieved context spans multiple abstraction levels, enabling Bridge-RAG to maintain a comprehensive contextual scope and further improve the answer accuracy.

To make the abstract bridge tree practical at scale, Bridge-RAG plugs in the Cuckoo Filter recently published in CFT-RAG~\citep{li2026cftrag} as its entity lookup component. Our contribution at this layer is not a new probabilistic data structure but its \textit{integration} into the abstract-tree retrieval path: each filter entry now resolves an entity to a list of abstract-node addresses, so a single filter hit suffices to recover the full entity-to-abstract-to-chunk path, with chunk addresses derived deterministically from the matched abstracts. For the underlying filter mechanics, we refer the reader to CFT-RAG~\citep{li2026cftrag}.

Experimental results demonstrate that Bridge-RAG consistently outperforms all baselines on both accuracy and retrieval speed. On MedQA, Bridge-RAG improves ROUGE-L by up to 9.8\% and BLEU by up to 11.6\% over Tree RAG, while being $1.7\times$ faster. These gains confirm that our \textit{hierarchical abstract bridge tree} effectively captures complex semantic relationships and context, while the integrated Cuckoo Filter from CFT-RAG~\citep{li2026cftrag} sustains high retrieval efficiency as the knowledge base scales.

\section{Related Work}
\label{sec:related}

\paragraph{Retrieval-Augmented Generation.}
RAG systems ground LLM outputs in external knowledge to reduce hallucination~\citep{patrick2020retrieval}.
The simplest variants embed document chunks into a vector store and retrieve the top-$k$ by cosine similarity~\citep{karpukhin2020dense}.
Because flat retrieval treats every chunk independently, it often returns semantically similar but contextually disjoint passages, leaving the generator to reconcile fragments on its own~\citep{gao2024retrievalaugmented}.
Subsequent work has scaled dense retrieval significantly: RETRO~\citep{borgeaud2022improving} retrieves from trillions of tokens at pre-training time, and Atlas~\citep{izacard2022few} demonstrates that coupling a retriever with a language model enables strong few-shot performance on knowledge-intensive tasks.
Despite these advances, the core limitation of flat retrieval persists: individual passages lack the broader semantic context needed for complex, multi-hop questions.

\paragraph{Structured Knowledge for RAG.}
To provide richer context, recent work organizes external knowledge into graphs or trees before retrieval.
Graph RAG~\citep{Darrin2024from,hu2024graggraphretrievalaugmentedgeneration} constructs entity--relation graphs and performs multi-hop traversal over node embeddings, effectively capturing relational context.
Tree-structured approaches such as T-RAG~\citep{fatehkia2024t} arrange entities hierarchically so that ancestor nodes supply broader context.
RAPTOR~\citep{raptor} builds a recursive abstraction tree by clustering and summarizing chunks bottom-up, enabling retrieval at multiple granularity levels.
Bridge-RAG shares RAPTOR's intuition that intermediate summaries improve retrieval quality, but differs in two key respects: (i)~our abstracts are deterministic fixed-window groupings rather than stochastic clusters, which makes the tree structure reproducible and index-friendly; and (ii)~we pair the tree with an $O(1)$ Cuckoo Filter entry point rather than relying on embedding-based search at every tree level.

\paragraph{Multi-Hop and Reasoning-Augmented Retrieval.}
Complex questions often require synthesizing evidence from multiple passages, motivating multi-hop retrieval designs.
IRCoT~\citep{trivedi2023interleaving} interleaves chain-of-thought reasoning with retrieval steps, using intermediate reasoning to formulate follow-up queries.
HippoRAG~\citep{gutierrez2024hipporag} models long-term memory inspired by hippocampal indexing theory, maintaining a knowledge graph that supports multi-hop associative retrieval.
Hyper-RAG~\citep{luo2026hypergraphrag} extends graph-based retrieval to hypergraphs for capturing higher-order entity relationships.
These approaches focus on the \emph{reasoning} side of multi-hop QA; Bridge-RAG complements them by providing a hierarchical \emph{index structure} that surfaces multi-level context in a single retrieval pass, without requiring iterative query reformulation.

\paragraph{Efficient Retrieval with Probabilistic Data Structures.}
Approximate membership structures such as Bloom filters~\citep{bloom1970space} and Cuckoo filters~\citep{fan2014cuckoo,pagh2001cuckoo} have long been used to accelerate set-membership queries.
CFT-RAG~\citep{li2026cftrag} is the first to embed a Cuckoo Filter into a RAG pipeline, storing entity-to-tree-node mappings for constant-time lookup.
Bridge-RAG builds on this line of work but shifts the contribution focus: whereas CFT-RAG uses the filter to locate \emph{entities} in a flat entity tree, we use it to locate \emph{abstracts}---semantic summary nodes that bridge entities and document chunks.
The filter itself is adopted from CFT-RAG without modification; our novelty lies in the hierarchical abstract bridge tree and the multi-level context retrieval strategy that operates over it.
In summary, the core distinction of Bridge-RAG is the \textit{abstract bridge} layer: by interposing semantic summaries between entities and raw chunks, we convert a flat entity lookup into a structured, multi-granularity retrieval path that neither flat-filter nor graph-traversal approaches provide.

\section{Methodology}
In this section, we present Bridge-RAG. Its centerpiece is a hierarchical abstract bridge tree that improves retrieval accuracy by bridging query entities, abstracts, and document chunks. To keep the retrieval over this hierarchy efficient at scale, we further integrate the Cuckoo Filter component from CFT-RAG~\citep{li2026cftrag} for $O(1)$ entity-to-abstract lookup.

\subsection{Hierarchical Abstract Tree Structure}
\label{subsec:abstract_tree}
The foundation of our approach lies in the hierarchical abstract tree structure, which addresses the accuracy limitations of traditional RAG systems. Abstracts are built by semantically grouping $n$ consecutive document chunks into higher-level knowledge units, where $n$ is a configurable window size (set to 5 in our experiments). Each abstract receives a pair identifier $i$ and subsumes chunk IDs $n_i, n_{i{+}1}, \ldots, n_{i{+}n{-}1}$. The $n$ constituent chunks are merged via embedding aggregation or summarization to produce a concise semantic summary, enabling efficient abstraction-layer retrieval.

Abstracts are organized hierarchically into tree structures, where upper-level abstracts represent more general and abstract concepts, while lower-level abstracts capture more specific and detailed information. This hierarchical organization enables multi-level semantic retrieval: when a query entity is identified, we not only retrieve the corresponding abstract and its associated document chunks, but also trace the abstract's parent-child relationships in the tree. Formally, given an initial abstract set $\mathcal{A}_0 = \{a_i\}$ retrieved from the Cuckoo Filter, we expand it through hierarchical traversal:
\begin{equation}
\mathcal{A}_{\text{expanded}} = \mathcal{A}_0 \cup \bigcup_{a \in \mathcal{A}_0} \left( \mathcal{P}_d(a) \cup \mathcal{C}_d(a) \right)
\end{equation}
where $\mathcal{P}_d(a)$ denotes the set of parent abstracts up to $d$ levels upward, and $\mathcal{C}_d(a)$ denotes the set of child abstracts up to $d$ levels downward. By retrieving these related abstracts along with their corresponding document chunks, we construct a comprehensive context spanning multiple abstraction levels, significantly improving answer accuracy by providing both high-level semantic understanding and granular textual details.

\subsection{Efficient Lookup via Cuckoo Filter}
\label{subsec:cuckoo_filter}
Traversing the abstract tree to locate the abstracts that contain a given entity introduces non-trivial overhead. To remove this bottleneck without re-inventing the wheel, we incorporate the Cuckoo Filter component in CFT-RAG~\citep{li2026cftrag}, which performs constant-time entity lookup and is well-suited as the entry point of our entity-to-abstract pathway.

\begin{figure*}[t]

    \centering
    \includegraphics[width=\textwidth]{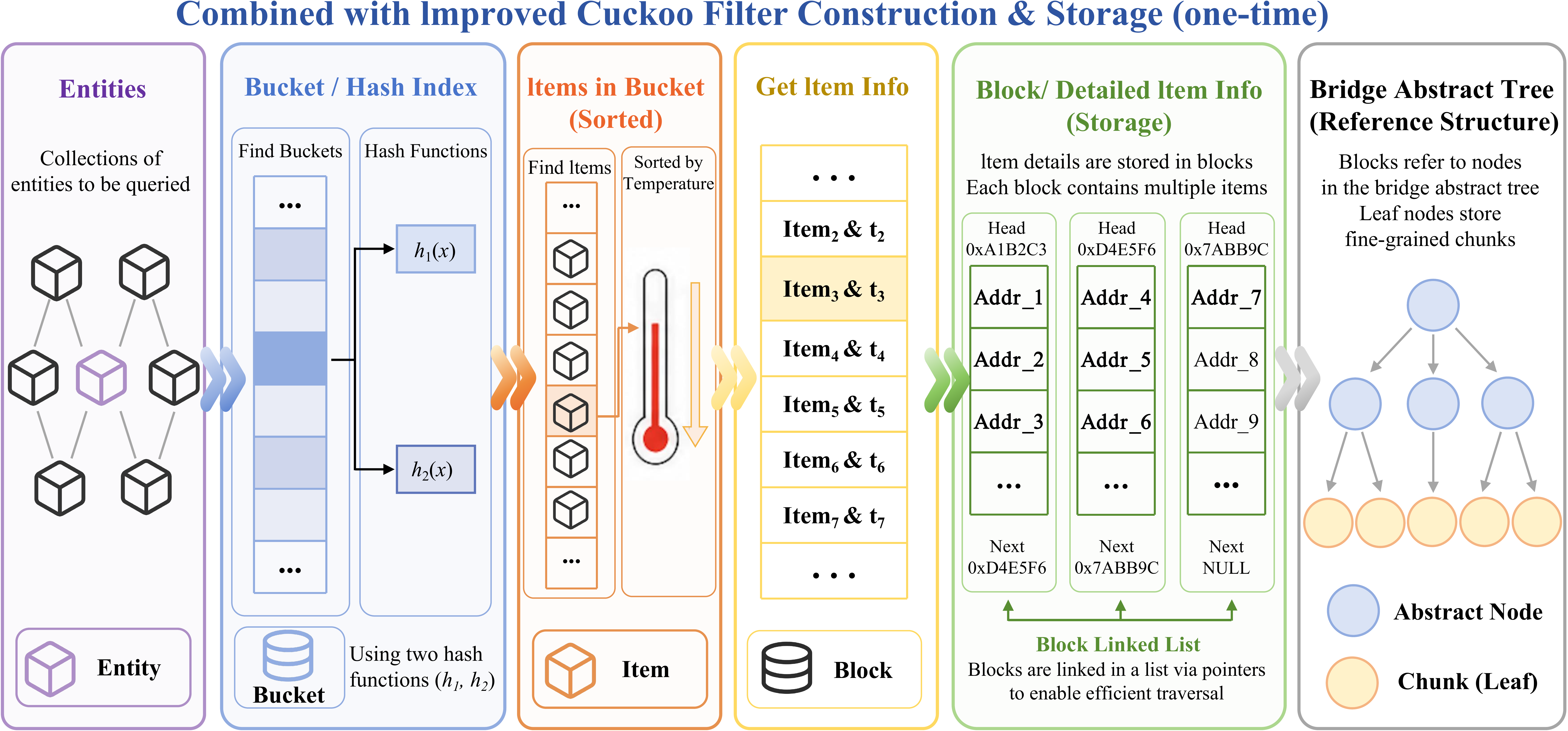}
        \caption{Detailed workflow of Bridge-RAG. Entities recognized from the query are looked up via the Cuckoo Filter in $O(1)$ time. For each matched abstract, the hierarchical context enrichment module traverses parent and child nodes in the abstract tree. The retrieved multi-level context (target, parent, and child abstracts with their chunks) is integrated into a comprehensive prompt for the LLM.}
    \label{fig:my_label2}
\end{figure*}

Our integration only touches the filter at the interface level: each filter entry, instead of pointing to a flat entity or a raw chunk address, is bound to the abstract-node addresses associated with the entity. Because every abstract with pair\_id $i$ deterministically owns chunks $n_i$ through $n_{i{+}n{-}1}$, the document-chunk addresses are recoverable from the matched abstracts, so a single filter hit suffices to expose the full entity-to-abstract-to-chunk path. The resulting retrieval workflow is illustrated in Figure~\ref{fig:my_label2}; the internal mechanics of the filter (fingerprint hashing, bucket layout, eviction, and access-frequency reordering) are exactly those of CFT-RAG~\citep{li2026cftrag} and are summarized in Appendix~\ref{sec:appendix_filter} for completeness.

\subsection{Context Generation with Entity-Abstract-Chunk Bridge}
\label{subsec:context_generation}
Our context generation process leverages the hierarchical abstract tree together with the integrated Cuckoo Filter to achieve both high accuracy and efficiency. The workflow demonstrates a bridging mechanism that connects query entities to abstracts and then to specific document chunks, where abstracts serve as the crucial intermediate layer.

The workflow consists of five steps: (1) \textbf{Entity Recognition}: Key entities are identified using entity recognition techniques (e.g., SpaCy). (2) \textbf{Entity-to-Abstract Lookup}: For each recognized entity, the integrated Cuckoo Filter from CFT-RAG~\citep{li2026cftrag} is queried, returning the set of abstracts that contain it and updating its access counter. (3) \textbf{Hierarchical Context Enrichment}: We trace parent-child relationships to expand the abstract set, retrieving parent and child abstracts up to a configurable maximum depth $d$. (4) \textbf{Abstract-to-Chunk Bridge}: We retrieve all chunks associated with the expanded abstract set, where each abstract is linked to $n$ chunks. (5) \textbf{Similarity-based Selection}: We compute cosine similarity between query embedding $\mathbf{q}$ and chunk embeddings $\mathbf{c}_j$:
\begin{equation}
\text{sim}(\mathbf{q}, \mathbf{c}_j) = \frac{\mathbf{q} \cdot \mathbf{c}_j}{\|\mathbf{q}\| \|\mathbf{c}_j\|}
\end{equation}
We then select the top-$k$ chunks $\mathcal{C}_{\text{top-k}}$ with the highest similarity scores:
\begin{equation}
\mathcal{C}_{\text{top-k}} = \arg\max_{\mathcal{C}' \subseteq \mathcal{C}, |\mathcal{C}'| = k} \sum_{c_j \in \mathcal{C}'} \text{sim}(\mathbf{q}, \mathbf{c}_j)
\end{equation}
where $\mathcal{C}$ is the set of all chunks associated with the expanded abstract set. The top-$k$ chunks are then selected, and their corresponding abstracts are retrieved. Finally, this multi-level contextual information (selected chunks and abstracts) is combined with the system prompt and query to form the final prompt for the LLM. The complete workflow is illustrated in Figure~\ref{fig:my_label0}, and the detailed algorithm is presented in Algorithm~1.

\begin{algorithm}[t]
\caption{Bridge-RAG Context Generation}
\KwIn{Query $q$, Cuckoo Filter $\mathit{CF}$, abstract forest $\mathcal{F}$, chunks-per-abstract $n$, return count $k$, max depth $d$}
\KwOut{Context string $\mathit{ctx}$}
\BlankLine
\tcc{Step 1: Entity recognition and Cuckoo Filter lookup}
$\mathbf{q} \gets \textsc{Embed}(q)$\;
$E \gets \textsc{RecognizeEntities}(q)$\;
$\mathcal{A}_0 \gets \emptyset$\;
\ForEach{$e \in E$}{
  $\mathcal{A}_0 \gets \mathcal{A}_0 \cup \mathit{CF}.\textsc{Lookup}(e)$\;
  $\mathit{CF}.\textsc{UpdateTemp}(e)$\;
}
\BlankLine
\tcc{Step 2: Hierarchical context enrichment}
$\mathcal{A} \gets \mathcal{A}_0$\;
\ForEach{$a \in \mathcal{A}_0$}{
  $\mathcal{A} \gets \mathcal{A} \cup \mathcal{P}_d(a, \mathcal{F}) \cup \mathcal{C}_d(a, \mathcal{F})$\;
}
\BlankLine
\tcc{Step 3: Abstract-to-chunk bridge and ranking}
$\mathcal{C}_{\text{all}} \gets \bigcup_{a \in \mathcal{A}} \{n_{a + j} \mid 0 \le j < n \}$\;
$\mathcal{C}_k \gets \underset{S \subseteq \mathcal{C}_{\text{all}},\, |S|=k}{\arg\max} \sum_{c \in S} \text{sim}(\mathbf{q}, \textsc{Embed}(c))$\;
\BlankLine
\tcc{Step 4: Prompt construction}
$\mathit{ctx} \gets \textsc{BuildPrompt}(\mathcal{C}_k, \mathcal{A}, q)$\;
\Return{$\mathit{ctx}$}\;
\end{algorithm}

\section{Experiments}
\label{sec:experiments}

This section reports the empirical study of Bridge-RAG. We first list the baselines and datasets, then present the main accuracy and retrieval-latency numbers, and finally analyze the effect of the hierarchical depth $d$ and the locality-aware bucket reordering. Our goal is to understand whether the abstract bridge tree pays off in answer quality, and whether the resulting retrieval pipeline stays fast at scale.

\subsection{Baseline}
We compare Bridge-RAG against three representative RAG variants that span the no-structure, graph-structured, and tree-structured ends of the design space, so that the contribution of our abstract bridge tree can be attributed cleanly.

\textbf{Naive RAG}
This baseline follows the standard retrieve-then-generate pipeline without any hierarchical or filtering enhancement. The top-K chunks returned by cosine similarity are directly concatenated with the user query and fed to the LLM. No abstract layer, graph structure, or approximate acceleration is employed, providing a clean reference for measuring the gains introduced by later optimizations.

\textbf{Graph RAG}
A graph-based variant in which entities and their typed relations form a directed graph; retrieval is accelerated with approximate nearest neighbor (ANN) indexing over node embeddings, and multi-hop expansion along the graph provides the surrounding context that the generator conditions on. This baseline isolates the gains of graph-style relational expansion from those of hierarchical abstraction.

\textbf{Tree RAG}
An entity-tree variant that uses approximate nearest neighbor (ANN) indices such as FAISS or HNSW in place of exact similarity search, returning top-$K$ candidates per query. This baseline isolates the gains attributable to our abstract bridge tree (which replaces flat entity leaves with multi-level abstract nodes) from those attributable to the underlying tree structure itself.

\textbf{RAPTOR}~\citep{raptor}
A recursive abstractive processing method that clusters document chunks bottom-up and summarizes each cluster to form a tree. At retrieval time, RAPTOR searches across all tree levels by embedding similarity. Unlike Bridge-RAG, RAPTOR relies on stochastic clustering (rather than deterministic grouping) and uses embedding search at every level (rather than $O(1)$ filter lookup), making it less predictable and slower on large corpora.

\subsection{Bridge-RAG}
Bridge-RAG combines our hierarchical abstract bridge tree for multi-level context retrieval (Section~\ref{subsec:abstract_tree}) with the Cuckoo Filter component from CFT-RAG~\citep{li2026cftrag} for $O(1)$ entity-to-abstract lookup (Section~\ref{subsec:cuckoo_filter}). The implementation details are provided in the Methodology section.

\subsection{Datasets and Abstract Forest}
We evaluate on a long-context dataset, MedQA~\citep{med}, and a medium-sized dataset, AALCR~\citep{artificialanalysis2025aalcr}. From each corpus we run the pipeline of Section~\ref{subsec:abstract_tree} to obtain concepts, relations, and a filtered hierarchy; we then group every $n$ consecutive document chunks into one abstract node ($n{=}5$), which gives us an abstract forest whose leaves expose document content and whose internal nodes expose semantic context. This forest is the data that all Bridge-RAG retrieval queries are executed against.

\subsection{Setup}
The RAG orchestration code is written in Python, with the latency-sensitive components (the Cuckoo Filter and the forest traversal) implemented in C++. All measurements are taken on a NVIDIA RTX 4090 machine, and each configuration is run 100 times so that the reported numbers are averages over independent trials. We use Doubao Seed~1.6~\citep{doubao2024} as the generator, and judge answer quality with the standard suite of ROUGE, BLEU and BERTScore. 

\begin{table*}[htbp]
\centering

\resizebox{\textwidth}{!}{%
\begin{tabular}{llccccccc}
\hline
\textbf{Dataset} & \textbf{Method} & \textbf{ROUGE-1} & \textbf{ROUGE-2} & \textbf{ROUGE-L} & \textbf{BLEU} & \textbf{BERTScore} &  \textbf{Retrieval Time (s)}  \\
\hline
\multirow{6}{*}{\textbf{AALCR}}
 & Naive RAG  & 0.0890 & 0.0303 & 0.0812 & 0.0178 & 0.8244  & -  \\
 & Raptor~\citep{raptor} & 0.1294 & \textbf{0.0533} & 0.1253 & 0.0381 & 0.8279  & - \\
 & Tree RAG      & 0.1339 & 0.0501 & 0.1282 & 0.0298 & 0.8326  & 9.6377  \\
 & Graph RAG    & 0.1307 & 0.0344 & 0.1259 & 0.0262 & 0.8332  & 10.2088  \\
 & Bridge RAG(Depth=1)      & 0.1214 & 0.0393 & 0.1160 & 0.0390 & 0.8325 & \textbf{5.7162}  \\
 & Bridge RAG(Depth=2)      & 0.1437 & 0.0527 & 0.1365 & 0.0364 & \textbf{0.8402}  & 5.7231  \\
 & Bridge RAG(Depth=3)     & \textbf{0.1522} & 0.0495 & \textbf{0.1451} & \textbf{0.0429} & 0.8386  & 5.7564  \\
\hline
\multirow{6}{*}{\textbf{MedQA
}}
 & Naive RAG & 0.3081 & 0.1529 & 0.3048 & 0.0679 & 0.8632  & - \\
 & Raptor~\citep{raptor} & 0.3382 & 0.1452 & 0.3225 & 0.0758 & 0.8672 & - \\
 & Tree RAG      & 0.3498 & 0.1415 & 0.3466 & 0.0733 & 0.8698 & 8.7238   \\
 & Graph RAG    & 0.3533 & 0.1412 & 0.3478 & 0.0698 & 0.8712 & 9.6782   \\
 & Bridge RAG(Depth=1)      & 0.3501 & 0.1467 & 0.3452 & 0.0776 & 0.8741  & \textbf{5.0237}  \\
 & Bridge RAG(Depth=2)      & 0.3599 & 0.1430 & 0.3542 & 0.0722 & 0.8772  & 5.0354  \\
 & Bridge RAG(Depth=3)    & \textbf{0.3884} & \textbf{0.1678} & \textbf{0.3805} & \textbf{0.0818} & \textbf{0.8784}  & 5.0618  \\
\hline
\end{tabular}%

}
\caption{Main results on the benchmarks. Bold marks the best performance per dataset.}
\label{tab:combined_evaluation}
\end{table*}

\subsection{Comparison Experiment}
For MedQA we sample 1{,}000 questions uniformly while 100 questions for AALCR and we report the per-query retrieval latency together with answer-quality scores (ROUGE, BLEU, BERTScore) averaged over the sample. Headline numbers for MedQA and AALCR are collected in Table~\ref{tab:combined_evaluation}.

Bridge-RAG achieves the best performance across both datasets. On AALCR, Bridge-RAG (Depth=3) improves BLEU by 44.0\% over Tree RAG while being $1.67\times$ faster, and by 63.7\% over Graph RAG while being $1.77\times$ faster. Compared to Naive RAG, Bridge-RAG improves ROUGE-L by 78.7\% and BLEU by 141.0\%.

On MedQA, Bridge-RAG improves ROUGE-L by 9.8\% and BLEU by 11.6\% over Tree RAG, while being $1.72\times$ faster. Compared to Graph RAG, Bridge-RAG improves ROUGE-L by 9.4\% and BLEU by 17.2\%, while being $1.91\times$ faster. Compared to Naive RAG, Bridge-RAG improves ROUGE-L by 24.8\% and BLEU by 20.5\%.

Among all baselines, RAPTOR~\citep{raptor} is the most architecturally similar to Bridge-RAG, as both build hierarchical summary structures over document chunks. However, the two systems differ fundamentally in their design philosophy. RAPTOR uses unsupervised clustering (e.g., Gaussian Mixture Models) to group chunks, producing clusters of variable size that depend on the clustering hyperparameters and random initialization. Bridge-RAG instead uses deterministic fixed-window grouping (every 5 consecutive chunks), which guarantees reproducibility and enables the Cuckoo Filter's direct address mapping. Empirically, RAPTOR underperforms Bridge-RAG on both datasets (Table~\ref{tab:combined_evaluation}). On MedQA, Bridge-RAG (Depth=3) outperforms RAPTOR on all metrics, with a 18.0\% gain in ROUGE-L and 7.9\% in BLEU. On AALCR, the gap narrows on ROUGE-2 (where RAPTOR's clustering occasionally captures useful cross-chunk correlations), but Bridge-RAG maintains clear leads on ROUGE-1, ROUGE-L, BLEU, and BERTScore.

The speed advantage comes from the integrated Cuckoo Filter's direct entity lookup ($O(1)$ complexity), compared to Tree-RAG's exhaustive BFS traversal. Moreover, our method only needs to calculate the similarity between the query and the candidate chunk pool, rather than the full set of chunks, which greatly speeds up the process. The accuracy improvement stems from Bridge-RAG's explicit hierarchical abstraction levels (parent, target, child) that capture multi-level semantic relationships, providing more global and semantically coherent context than graph-based traversal or direct chunk retrieval. The multi-level context retrieval strategy traces parent-child relationships in the abstract tree, retrieving context from multiple abstraction levels to enable more accurate and contextually coherent answers.

Comparing different depth settings reveals a trade-off between accuracy and efficiency. The accuracy improvement with increasing depth comes from retrieving context from more abstraction levels: Depth=2 and Depth=3 additionally retrieve deeper parent and child abstracts, providing broader semantic context and more granular details. The increased retrieval time is due to the additional tree traversal operations required to access parent and child nodes, but remains efficient thanks to the fast entity-to-abstract mapping that minimizes the overhead of hierarchical expansion.

\subsection{Complexity Analysis}
\label{subsec:complexity}

\paragraph{Time complexity.}
Given a query with $|E|$ recognized entities, Bridge-RAG performs (i)~$|E|$ Cuckoo Filter lookups, each costing $O(1)$; (ii)~hierarchical expansion to depth $d$, visiting at most $O(b^d)$ abstract nodes per initial hit, where $b$ is the branching factor; and (iii)~a top-$k$ similarity selection over the candidate chunk pool $|\mathcal{C}|$, costing $O(|\mathcal{C}| \cdot \dim)$ where $\dim$ is the embedding dimension. In practice $b \leq 5$ and $d \leq 3$, so the expansion is bounded by a small constant. The dominant cost is the similarity computation, which operates over a filtered candidate set rather than the full corpus---typically 2--3 orders of magnitude smaller. By contrast, Tree RAG performs BFS over the entire entity tree ($O(N)$ for $N$ entities), and Graph RAG traverses multi-hop neighborhoods that grow combinatorially with hop count.

\paragraph{Space complexity.}
The Cuckoo Filter stores one 12-bit fingerprint, one temperature counter, and one block-list pointer per entity, yielding $O(|E_{\text{total}}|)$ space with a small per-entry constant ($\sim$4 bytes). The abstract forest adds $O(|A|)$ nodes. Overall, Bridge-RAG's indexing overhead is linear in corpus size and negligible compared to the embedding store.

\section{Discussion}
\label{subsec:discussion}

\paragraph{Why abstracts outperform flat entities.}
A key insight from our results is that the abstract layer provides a semantic granularity that neither raw chunks nor flat entities offer. Raw chunks are too fine-grained: they capture local details but miss the broader context needed for multi-hop questions. Flat entities, as used in standard Tree RAG, are too coarse: they index individual concepts without encoding how those concepts relate within a passage. Abstracts occupy a middle ground---each one summarizes a coherent group of chunks and naturally encodes local relationships, making them effective bridges for the entity-to-chunk retrieval path.

\paragraph{Depth selection in practice.}
Our depth study reveals that Depth=3 consistently achieves the best accuracy, but the marginal gain from Depth=2 to Depth=3 is smaller than from Depth=1 to Depth=2 (Table~\ref{tab:combined_evaluation}). This diminishing return suggests that most useful context resides within two levels of the target abstract. For latency-sensitive deployments, Depth=2 offers a practical sweet spot---retaining most of the accuracy benefit at lower traversal cost. Adaptive depth selection, where the system chooses depth based on query complexity, is a promising direction we leave for future work.

\paragraph{Interaction between the abstract tree and the Cuckoo Filter.}
While Bridge-RAG adopts the Cuckoo Filter from CFT-RAG without modification, the two components are not merely stacked independently. The abstract tree changes the semantics of what the filter indexes: instead of mapping entities to tree nodes in a flat entity hierarchy, each filter entry now maps to an abstract node that deterministically owns chunks and participates in a parent--child hierarchy. This means a single filter hit not only locates the target content but also opens a path to hierarchically related context---an affordance that a flat entity-to-chunk mapping cannot provide.

\paragraph{Cross-dataset generalization.}
Bridge-RAG's gains are consistent across MedQA (a domain-specific medical corpus with specialized terminology) and AALCR (a general long-context reasoning benchmark), suggesting that the abstract bridge tree is not tuned to a particular domain. The hierarchical structure is especially beneficial for MedQA, where medical concepts form natural taxonomies (e.g., diseases $\to$ symptoms $\to$ treatments) that align well with the parent--child organization of the abstract tree. On AALCR, the improvement is driven more by the multi-level context providing broader reasoning scope for complex, multi-hop questions. This cross-domain consistency suggests that the abstract bridge tree captures a general structural pattern rather than domain-specific regularities.

\paragraph{Scalability.}
Bridge-RAG's $O(1)$ filter lookup is independent of corpus size, whereas Tree RAG's BFS scales linearly and Graph RAG's multi-hop expansion grows combinatorially. Only the final top-$k$ selection over candidate chunks grows with the corpus, and this set is bounded by the small number of matched abstracts. Our $1.7\times$--$1.9\times$ speedups over Tree RAG and Graph RAG on AALCR and MedQA should therefore widen further on larger corpora where traversal costs dominate.

\paragraph{When does Bridge-RAG underperform?}
Bridge-RAG does not dominate on every metric. On AALCR, RAPTOR achieves higher ROUGE-2 at Depth=1 and Depth=2 (Table~\ref{tab:combined_evaluation}), suggesting that RAPTOR's clustering occasionally captures bigram-level patterns that fixed-window grouping misses. Bridge-RAG at Depth=1 also underperforms Graph RAG on BERTScore for AALCR, confirming that the abstract tree's advantage requires multi-level traversal---the bridge layer alone, without parent--child expansion, provides limited benefit.

\section{Conclusion}
We presented Bridge-RAG, a retrieval-augmented generation framework whose core contribution is a \textit{hierarchical abstract bridge tree} that connects query entities to document chunks via an intermediate abstract layer, retrieving coherent multi-level context through parent--child traversal. To keep this hierarchical retrieval fast at scale, we delegate the entity lookup step to the Cuckoo Filter of CFT-RAG~\citep{li2026cftrag}, which integrates seamlessly with our pipeline. Experiments on MedQA and AALCR show consistent accuracy gains over all baselines (up to 9.8\% ROUGE-L and 44\% BLEU over structured RAG methods), with $1.7\times$--$1.9\times$ faster retrieval than Tree RAG and Graph RAG. Our analysis further shows that the abstract layer provides a semantic granularity unavailable in flat entity or chunk-based retrieval, and that hierarchical depth of 2--3 captures most useful context with bounded overhead. These results demonstrate that bridging the gap between entities and chunks via learned summaries is a promising paradigm for scalable, high-quality RAG.

Looking ahead, we identify three promising directions. First, replacing the fixed five-chunk window with content-aware segmentation could better align abstract boundaries with natural topic shifts. Second, adaptive depth selection---choosing the traversal depth based on query complexity or entity density---could further optimize the accuracy--latency trade-off for heterogeneous workloads. Finally, combining the hierarchical abstract bridge with iterative reasoning strategies such as chain-of-thought retrieval~\citep{trivedi2023interleaving} could unlock stronger performance on multi-hop questions that require sequential evidence gathering across distant parts of the corpus.

\section*{Limitations}
Bridge-RAG assumes a fixed five-chunk grouping window for abstract construction. While this design enables deterministic indexing and efficient Cuckoo Filter integration, it may not generalize well to all document structures---corpora with highly variable paragraph lengths or non-linear layouts (e.g., tables, dialogue transcripts) could benefit from content-aware or variable-length segmentation. A systematic study of how grouping size affects the accuracy--efficiency frontier is left for future work. Moreover, although the accuracy has been improved, the construction cost of the bridge forest is very high. Reducing both the time and token cost is also included in our future work. Finally, Bridge-RAG inherits the hallucination risk of the frozen backbone LLM: when retrieved chunks themselves contain conflicting or outdated facts, the system can propagate rather than correct the error.

\bibliography{custom}

\clearpage
\newpage
\appendix
\section{Cuckoo Filter Details}
\label{sec:appendix_filter}

Bridge-RAG adopts the Cuckoo Filter from CFT-RAG~\citep{li2026cftrag} without modification to its internal mechanics. We summarize the key aspects for self-containedness.

\paragraph{Entity Insertion.}
Each entity is represented by a short fingerprint $f(x)$. Two candidate buckets $i_1 = h(x)$ and $i_2 = i_1 \oplus h(f(x))$ are computed. If either bucket has a free slot, the fingerprint, an initial temperature counter, and the entity's abstract-address list are stored there. Otherwise, the standard cuckoo eviction routine displaces an existing entry, subject to a fixed kick budget.

\paragraph{Access-Frequency Reordering.}
Every entity carries a temperature counter $T(e)$ incremented on each successful lookup. The filter periodically sorts entries within each bucket in descending order of $T(e)$, so that frequently accessed entities are found first during the linear per-bucket scan.

\paragraph{Storage Layout.}
Each bucket slot holds: (i)~a 12-bit fingerprint, (ii)~the temperature counter at the head of the address chain, and (iii)~a pointer to a block linked list whose nodes store abstract pair\_ids. From each pair\_id $i$, chunks $n_i$ through $n_{i{+}n{-}1}$ are derived deterministically. The resulting layout is shown in Figure~\ref{fig:my_label2}.

\section{Evaluation Details}
\subsection{Prompt Template}
\begin{promptbox}{Bridge-RAG Prompt Template}
\small\sffamily
\textbf{System:} Answer the question using the provided information.\\[0.8em]
\textbf{Information:}\\
\texttt{<retrieved chunks from top-k similarity search>}\\[0.8em]
\textbf{Abstracts:}\\
\texttt{<hierarchical abstract summaries>}\\[0.8em]
\textbf{Question:}\\
\texttt{<user query>}\\[0.8em]
\textbf{Assistant:}\\
\texttt{<generated answer>}
\end{promptbox}

For baseline RAG (without abstract tree), the prompt template is simplified to:

\begin{promptbox}{Baseline RAG Prompt Template}
\small\sffamily
\textbf{System:} Answer the question using the provided information.\\[0.8em]
\textbf{Information:}\\
\texttt{<retrieved chunks from vector similarity search>}\\[0.8em]
\textbf{Question:}\\
\texttt{<user query>}\\[0.8em]
\textbf{Assistant:}\\
\texttt{<generated answer>}
\end{promptbox}

\vspace{1cm}

\section{The Use of LLMs}
An LLM was used solely as a writing aid during paper preparation, to flag grammar and spelling issues and to suggest sentence-level rewrites for fluency. All technical content, claims, and experimental results were authored and verified by the human authors.


\end{document}